\documentclass[aps,pra,twocolumn,showpacs]{revtex4}
\usepackage{graphicx}

\begin{document}

\title{Chaotic oscillation in attractive Bose-Einstein condensate under an
impulsive force}

\author{Paulsamy Muruganandam}
\affiliation{Instituto de F\'{\i}sica Te\'orica, Universidade Estadual
Paulista, 01.405-900 S\~ao Paulo, S\~ao Paulo, Brazil}

\author{Sadhan K. Adhikari}

\affiliation{Instituto de F\'{\i}sica Te\'orica, Universidade Estadual
Paulista, 01.405-900 S\~ao Paulo, S\~ao Paulo, Brazil}

\date{\today}
\begin{abstract}

For an attractive trapped Bose-Einstein condensate an imaginary three-body
recombination loss term and an imaginary linear source term are usually included
in the Gross-Pitaevskii (GP) equation for a proper account of dynamics. Under
the action of an impulsive force, generated by suddenly changing the atomic
interaction or the trapping potential, the solution of this complex GP equation
for attractive interaction is found to lead to a very long term chaotic
oscillation.

\end{abstract}

\pacs{03.75.Fi}

\maketitle

Since the successful detection \cite{1,ex2} of Bose-Einstein condensates (BEC)
in dilute bosonic atoms employing magnetic trap at ultra-low temperature, there
has been great theoretical and experimental interest in condensates with
attractive interaction \cite{ex2}.  For attractive interaction the condensate
is stable for a maximum critical number $N_{\mbox{cr}}$ of atoms \cite{ex2}. 
When the number of atoms increases beyond this critical number, due to
interatomic attraction the condensate collapses emitting atoms until the number
of atoms is reduced below $N_{\mbox{cr}}$ and a stable configuration is
reached. With a supply of atoms from an external source the condensate can grow
again and thus a series of collapses can take place, which was observed
experimentally in the BEC of $^7$Li with attractive interaction \cite{ex2}.
Theoretical analyses based on the mean-field Gross-Pitaevskii (GP) equation
\cite{8,11} also confirm the collapse \cite{11,th1,th2}.

In a recent classic experiment by the JILA group on the BEC of $^{85}$Rb, a
sustained harmonic oscillation of an attractive condensate had been observed
and measured when the scattering length of atomic interaction was suddenly
changed \cite{ex4}.  It is possible to manipulate the  scattering length  by an
external magnetic field via a Feshbach resonance \cite{fs,ex3}.   The sudden
change in scattering length or the harmonic oscillator trapping potential
constitute a general class of impulsive force in the GP equation which can be
realized experimentally. We show that under the action of an impulsive force
the solution of the GP equation for attractive interaction with a pentic loss
term and a linear source term can exhibit sustained chaotic oscillation.  The
imaginary pentic  term in the GP equation accounts for three-body recombination
loss and the imaginary linear term accounts for growth due to an external
source of atoms.

There have been several studies of chaotic dynamics in the nonlinear
Schr\"odinger \cite{nls},  Landau-Ginzburg\cite{lz}, and  the GP \cite{gp2,gam}
equation. But all these studies were motivated from a numerical or mathematical
point of view without phenomenological consequence. The chaotic oscillation we
study here may be observed in the laboratory by slightly manipulating the
strength of the linear source term responsible for supplying atoms to the
condensate. In the absence of any source term or for a weak  source term the
oscillation is periodic after the application of the impulsive force.  With a
small increase in the source term the motion turns out to be chaotic. The JILA
group has performed a classic experiment where they have been able to observe
and measure the frequency of oscillation of a condensate after the application
of an impulsive force generated by changing the scattering length to a negative
value \cite{ex4}. Such an experiment might be able to detect the transition
from periodic to chaotic oscillation of a condensate under the action of an
impulsive force.

We consider for our present study on the numerical solution \cite{sk1,sk3} of
the GP equation \cite{8} for a spherically symmetric harmonic trap.  In the GP
equation we include a pentic three-body nonlinear recombination term which
accounts for the decay of the strongly attractive condensate and a linear
source term. 

The GP equation in this case in dimensionless units can be written as
\cite{th1,th2}
\begin{eqnarray} \label{c}
\biggr[-\frac{\partial^2}
{\partial x^2} & +& \frac{x^2}{4}+ 2\sqrt 2 n\left|
\frac{\varphi(x,t)}{x}
\right| ^2- i\xi \left|\frac{\varphi(x,t)}{x}
\right| ^4 \nonumber \\ &+& i\gamma-i\frac{\partial
}{\partial t}\biggr] \varphi (x,t)=0,
\end{eqnarray}
where $n=Na/l$, $\gamma$ is the coefficient of the linear source term and $\xi$
is the coefficient of the pentic three-body recombination term. Here the
distance $x$, time $t$, and the  spherically symmetric wave function $\phi(x,t)
\equiv \varphi(x,t)/x$ are expressed in units of $l/\sqrt 2$,  $\omega ^{-1}$,
and $(\sqrt 2 \pi l^3)^{-1/2}$, where $l=\sqrt{(\hbar/m\omega)}$ and $\omega$
is the frequency of the harmonic oscillator trap,  $m$ is the mass of a single
atom, $N$ is the number of atoms in the condensate, and $a$ the scattering
length of atomic interaction. The normalization condition for the wave function
is
\begin{eqnarray}\label{n1}
\int_0^\infty dx |\varphi(x,t)|^2=1.
\end{eqnarray}
We solve Eq. (\ref{c}) above by the Crank-Nicholson time propagation \cite{sk1}
after discretization with $\gamma \ne 0$ and $\xi \ne 0$ starting from the
known harmonic oscillator solution for  $n=\xi=\gamma=0$. The $x$
discretization was performed with space step 0.1 up to a maximum $x$ of 40. The
time step was taken to be 0.01. The nonlinearity constant $n$ was increased by
steps of 0.0001 until the desired value is reached. Once the final nonlinearity
is reached the solution is then stabilized by iterating 100000 times which
corresponds to an interval of time $t=1000$.  This removes any transient
behavior in the solution which is important for a study of chaos. Then the wave
function is prepared for the simulation  of chaos after the application of the
impulsive force.  The oscillation of the system after the application of the
impulsive force  is best studied by considering the time evolution of the root
mean square (rms) radius $X$ and its time derivative $\dot X$. 

For attractive interaction a stable condensate can be formed for the minimum
value $-0.575$ for $n$ \cite{11,th1,th2}.  We consider the following three
numerical simulations for the study of chaos:  (i) On a preformed attractive
condensate with $n=-0.4$ we suddenly double the harmonic oscillator term from
$x^2/4$ to $x^2/2$ and study the resultant oscillation of the system for
different $\gamma$.  On a preformed repulsive condensate with $n=+0.4$ we
suddenly change the sign of the scattering length $a$ to (ii) $-a$ and also to
(iii) $-2a$ and study the resultant oscillation of the system for different
$\gamma$. In all simulations we take $\xi =0.0004$. In cases (i) and (ii) the
final value of $n(=-0.4>-0.575)$ permits stable condensate and in case (iii) 
final value of $n(=-0.8<-0.575)$ does not permit a stable solution of Eq.
(\ref{c}). However, the experiment of JILA \cite{ex4} has shown that in case
(iii) the number of particles can not decay immediately below the stability
limit and there could be sustained oscillation of the system with a value of
$n$ less than $-0.575$. We find that depending on the value of the source term
$\gamma$ there could be pronounced  chaos in all three cases above.

Once the wave function is prepared for the study of chaos as described above we
inflict the change corresponding to case (i), first, with $\gamma =0$. Only
periodic oscillation of rms radius $X$ and its derivative $\dot X$ is
observed.  With a slight increase in $\gamma $ to 0.0002, chaotic oscillation
is obtained. The change from periodic to chaotic motion is best illustrated by
plotting $\dot X$ versus $X$. For a periodic motion a closed loop appears in
the phase space plot of $\dot X$ versus $X$, whereas for a chaotic motion a
strange  attractor appears in the phase space plot. This is shown in Figs. 1
(a) and (b) for $\gamma = 0.0001$ (periodic) and 0.0002 (chaotic),
respectively,  for an interval of simulation time $t=200$ after inflicting the
impulsive force. With the increase of $\gamma$, the periodic motion of Fig.
1(a) changes to chaotic motion in Fig. 1(b). The present time is expressed in
units of $\omega^{-1}$. In a typical experimental situation the harmonic
oscillator trapping frequency $\nu \sim 50$ s$^{-1}$, and hence
$\omega^{-1}=(2\pi \nu)^{-1}\sim 0.005$ s.  Thus  the simulation time of
$t=200$ corresponds to 1 s which is inside the experimental observation period
of a typical set up \cite{ex4}. We continued the simulation till $t=20000$ and
the robust chaotic attractor seems to stay for ever, although it moves slowly
to a smaller value of $X$ with time. This is exhibited in Fig. 1 (c) where we
plot $\dot X$ versus $X$ for $0<t<20000$ which corresponds to an interval of
100 s. 
\begin{figure}[!h]
\begin{center}
\includegraphics[width=\columnwidth]{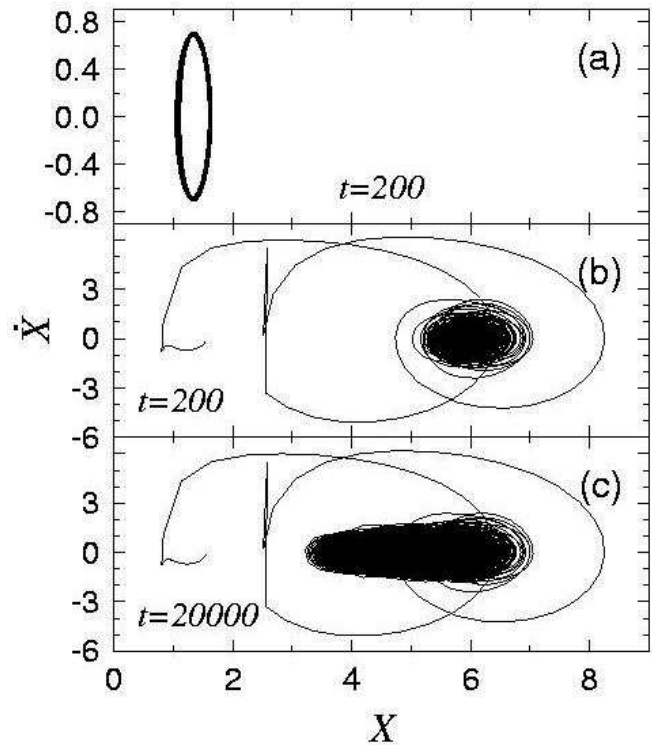}
\end{center}
\caption{$\dot X$ versus $X$ in case (i) when the harmonic oscillator potential
is suddenly doubled in Eq. (\ref{c}) for $\xi =0.0004$, $n=-0.4$ and (a)
$\gamma=0.0001$, $t=200$, (b)  $\gamma=0.0002$, $t=200$, and (c) 
$\gamma=0.0002$, $t=20000$.}
\end{figure}

Next we perform a similar simulation as  in Fig. 1 above for the case (ii)
where the sign of the scattering length is suddenly changed from positive to
negative exploiting a Feshbach resonance as in the experiment of JILA
\cite{ex4}. In simulation this corresponds to changing  $n$ from 0.4 to $-0.4$
(repulsive to attractive). The final $n (=-0.4 >-0.575)$ allows for a  stable
attractive condensate to be formed. Again for $\gamma =0$ and 0.0001 periodic
motions are  obtained  which change to chaotic motion as $\gamma$ is increased.
In Fig. 2 we plot $\dot X$ versus $X$ for $\gamma=0.0001$ and $\gamma=0.0002$ 
as in Fig. 1. In Fig. 2 (a) we show the periodic motion for $\gamma = 0.0001$
for time $t$ up to 200. Fully chaotic motion is obtained for $\gamma=0.0002$.
The chaotic motion for the initial interval of time 200 is shown in Fig. 2 (b)
and that for an interval of 20000 units of time is shown in Fig. 2 (c). The
robust chaotic attractor has started  to develop in Fig. 2 (b) and its
long-term evolution is shown in Fig. 2 (c). 
\begin{figure}[!ht]
\begin{center}
\includegraphics[width=\columnwidth]{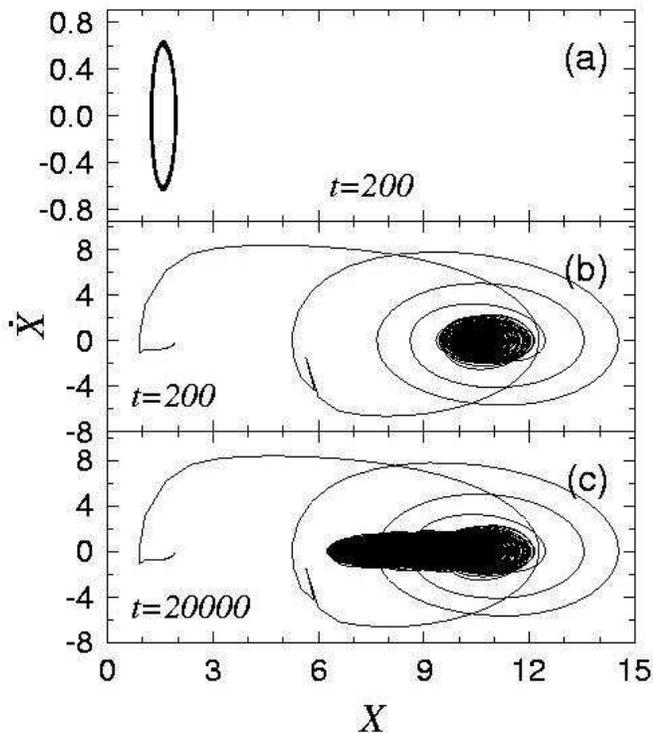}
\end{center}
\caption{$\dot X$ versus $X$ in case (ii) when the sign of the scattering
length $a$ was suddenly made negative from positive  in Eq. (\ref{c})   for
$\xi =0.0004$, $n=0.4$ and (a) $\gamma=0.0001$, $t=200$, (b)  $\gamma=0.0002$,
$t=200$, and (c)  $\gamma=0.0002$, $t=20000$.}
\end{figure}

Finally, we consider the case (iii) above. In this case the nonlinear term  $n$
has been suddenly changed from 0.4 to $-0.8$ (repulsive to attractive).
However, in this case the final $n (=-0.8 <-0.575)$ does not allow a stable
condensate to be formed. Such a strongly attractive condensate has been created
and observed in the laboratory in the experiment of JILA \cite{ex4}.   In this
case the condensate is unstable and due to interatomic attraction it starts to
shrink in size or collapse after the sudden change in the  nonlinear term.
Consequently, as the central density of the condensate increases, it starts to
emit particle through small explosions and tries to attain a more stable
configuration with a smaller number of particles \cite{ex4}. The condensate
exhibits oscillation during this process of collapse and explosion but it may
need a large amount of time before attaining the critical size with
$N_{\mbox{cr}}$ atoms.  No periodic oscillation was observed in this case even
for $\gamma =0$. In Fig. 3 (a) we plot $\dot X$ versus $X$ for the first 200
units of time for $\gamma = 0$, which shows a chaotic  attractor, which
continues to exist for any positive nonzero $\gamma$.  In Fig. 3 (b) we plot
the same for  $\gamma = 0.0002$, which shows the strange attractor for the
first 200 units of time. In Fig. 3 (c) we exhibit  the long-term  behavior of
this attractor for 20000 units of time. In the long term the attractor is more
spread in this case compared to Figs.  1 (c) and 2 (c). In case (iii) in the
long term the chaotic attractor moves to large $X$ whereas in cases (i) and
(ii) it moves to smaller $X$. 
\begin{figure}[!ht]
\begin{center}
\includegraphics[width=\columnwidth]{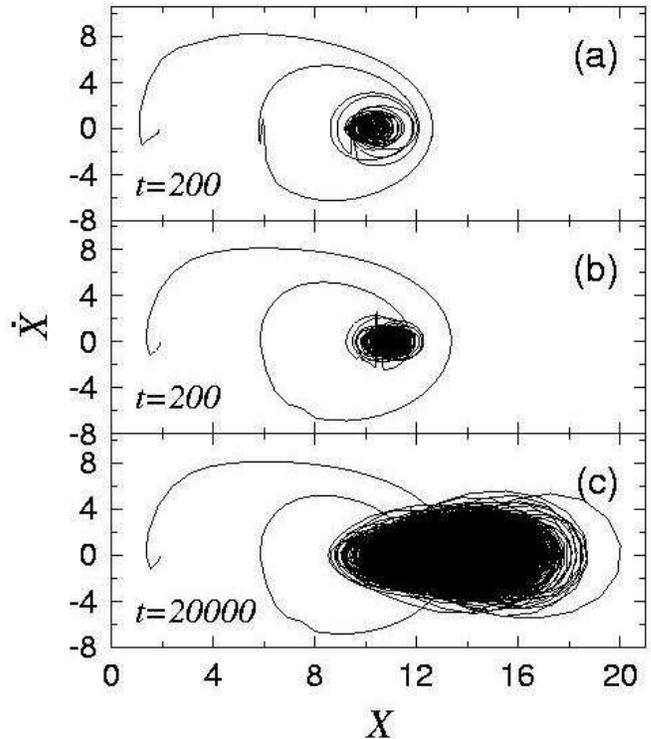}
\end{center}
\caption{$\dot X$ versus $X$ in case (iii) when the scattering length $a$ was
suddenly changed to $-2a$  in Eq. (\ref{c})  for $\xi =0.0004$, $n=0.4$ and (a)
$\gamma=0$, $t=200$, (b)  $\gamma=0.0002$, $t=200$, and (c)  $\gamma=0.0002$,
$t=20000$.}
\end{figure}
 
The presence of chaos is characterized by finding the Lyapunov exponents
\cite{kantz:1}. Since the quantities $X$ and $\dot X$ are not direct solutions
of the  nonlinear GP equation (\ref{c}) and are essentially the expectation
values, we use time series analysis to calculate the Lyapunov exponents
\cite{kantz:1}. In all the cases, we use the velocity-variable ($\dot X$) data
collected with the time interval $\delta t=0.1$. The reason to choose $\dot X$
is that it oscillates with time only around the steady average zero value,
whereas the time average of $X$ is not stationary and changes with time. From
one set of data for $\dot X(t)$ we construct other independent set(s) by
allowing a time lag $\tau$. The number of such independent sets constitute the
embedding dimension $m$ of the data sets for $\dot X(t)$ to be analyzed.  The
optimal time lag is found from the estimate of mutual information which shows
the independent nature of the reconstructed data sets \cite{fraser:1}. After
some experimentation we find that $m=3$ leads to reasonable values for all the
Lyapunov exponents.  All the calculations of the Lyapunov exponents are
performed with $m=3$ using the algorithm by Sano and Sawada \cite{sano:1}.

The calculation leads to three exponents in each of the cases (i), (ii), and
(iii) with $\gamma =0.0002$ shown in Figs. 1 (c), 2 (c), and 3 (c), 
respectively. In the case of (i), the time lag $\tau$ in the calculation of 
Lyapunov exponents was found to be 0.6 and in cases (ii), and (iii) it is found
to be 0.9. In Fig. 4 we plot the largest  Lyapunov exponent versus the time
used for the calculation  in the three cases above. In all three cases the
largest  exponent is found to lead to a  convergent  positive value at large
finite   time  which shows the existence of chaos in all the cases . We also
found that the oscillation in Fig. 3 (a) corresponding to case (iii) with
$\gamma = 0$ is also chaotic.
\begin{figure}[!ht]
\begin{center}
\includegraphics[width=\columnwidth]{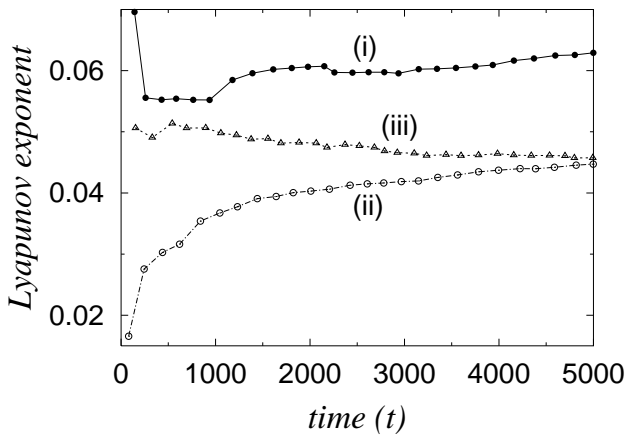}
\end{center}
\caption{The Lyapunov exponents for the chaotic attractors of Figs. 1 (c), 2
(c), and 3 (c) labeled (i), (ii), and (iii), respectively.}
\end{figure}

There has been confirmation of chaos in the nonlinear Schr\"odinger equation
\cite{nls} and in the Landau-Ginzburg equation \cite{lz} with similar nonlinear
terms as in the GP equation, although the details of the two equations are
different. Also, there has been theoretical prediction of chaos in a coupled
set of GP equations \cite{gp2}. The numerical study of chaos in  a collapsing
Bose-condensed gas in Ref. \cite{gam} is worth mentioning. In that work chaos
was confirmed in the normal time evolution of the GP equation (\ref{c}) without
any external impulsive force. The chaotic attractor in that study  was found to
stay only for a short interval of time ($\sim 500$), where in the present study
with impulsive forces the chaotic oscillations  are found to appear immediately
after the application of the impulsive force and stay for more than 20000 units
of time $t$.  This makes the analysis of the chaos via Lyapunov exponent more
reliable. 

The appearance  of chaos in nonlinear dynamics is of interest from a
theoretical point of view.  Here we have demonstrated the chaotic oscillation
in an attractive  BEC  under an impulsive force using the mean-field GP
equation. The use of the GP  equation in this study is  justified as  this
equation produces a faithful representation of the BEC for both  repulsive and
attractive interactions \cite{th1,th2}. We also performed simulation when the
condensate is repulsive after the application of the impulsive force. No
chaotic  oscillation was detected in that case. From this it seems that the
final-state interatomic attraction plays an important role in the  generation
of this chaotic dynamics.

In summary, from a numerical simulation based on the solution of the
Gross-Pitaevskii equation  (\ref{c}) for attractive interaction with an
absorptive pentic three-body recombination term ($\xi$) and a linear source
term $\gamma$, we find that sustained chaotic oscillation can result in a BEC
under the action of an impulsive force generated by suddenly changing the
interatomic scattering length or the harmonic oscillator trapping potential.

\acknowledgements
The work is supported in part by the Conselho Nacional de Desenvolvimento
Cient\'\i fico e Tecnol\'ogico and Funda\c c\~ao de Amparo \`a Pesquisa do
Estado de S\~ao Paulo of Brazil.

\end{document}